\documentclass[amssymb,nobibnotes,superscriptaddress,aps,notitlepage,twocolumn]{revtex4-1}

\usepackage{enumerate,amsmath}
\usepackage{graphicx}
\usepackage[dvips]{color} 

\newcommand{\red}{\color{black}} %

\newcommand{\dianorm}[1]{\| #1 \|_{\diamond}}
\newcommand{\Tr}{{\rm Tr}}

\newtheorem{theo}{Theorem}

\newtheorem{lemm}{Lemma}

\newtheorem{coro}{Corollary}

\begin{document}

\title{
Computational quantum-classical boundary
of noisy commuting quantum circuits
}%

\author{Keisuke Fujii}
\email{fujii.keisuke.2s@kyoto-u.ac.jp}
\address{The Hakubi Center for Advanced Research, Kyoto University, Yoshida-Ushinomiya-cho, Sakyo-ku, Kyoto 606-8302, Japan}
\address{Department of Physics, Graduate School of Science,
Kyoto University, Kitashirakawa Oiwake-cho,
Sakyo-ku, Kyoto 606-8502, Japan}
\address{Graduate School of Informatics, Kyoto University, Yoshida Honmachi, Sakyo-ku, Kyoto 606-8501, Japan}
\author{Shuhei Tamate}
\address{RIKEN Center for Emergent Matter Science, Wako, Saitama 351-0198, Japan}
\address{National Institute of Informatics, Hitotsubashi 2-1-2, Chiyoda-ku, Tokyo 101-8403, Japan}

\date{\today}
\begin{abstract}
It is often said that the transition from quantum to classical worlds
is caused by decoherence originated from an interaction between a system of 
interest and its surrounding environment.
Here we establish a computational quantum-classical boundary 
from the viewpoint of 
classical simulatability of a quantum system under decoherence.
Specifically, we consider commuting quantum circuits
being subject to decoherence.
Or equivalently, we can regard them as 
measurement-based quantum computation 
on decohered weighted graph states.
To show intractability of classical simulation in the quantum side,
we utilize the postselection argument 
and crucially strengthen it by taking noise effect into account.
Classical simulatability in the classical side
is also shown constructively 
by using both separable criteria in 
a projected-entangled-pair-state picture 
and the Gottesman-Knill theorem for mixed state Clifford circuits.
We found that when each qubit is subject to
a single-qubit complete-positive-trace-preserving noise,
the computational quantum-classical boundary is sharply given by 
the noise rate required for the
distillability of a magic state.
The obtained quantum-classical boundary of noisy quantum dynamics
reveals a complexity landscape of controlled quantum systems.
This paves a way to an experimentally feasible verification 
of quantum mechanics in a high complexity limit
beyond classically simulatable region.
\end{abstract}
\maketitle

\section*{Introduction}
Understanding a boundary between quantum and classical 
worlds is one of the most important quests in physics.
Sometimes it is said that 
decoherence originated from 
an interaction with an environment 
causes the transition from quantum to classical worlds~\cite{Zurek,ZurekRMP}.
However, the definition of ``quantumness"
varies depending on a situation where the system is located and 
a purpose of its usage.

One of the most successful definition would be
a violation of the Bell inequality~\cite{Bell};
if the measurement outcomes of Alice and Bob
violate the Bell inequality, 
the measurement outcomes cannot be expressed by any local hidden variable theory.
In this sense, whether or not the system obeys the Bell inequality 
serves as a quantum-classical boundary.
Nonlocality, or more widely, entanglement,
beyond the classical regime
is also utilized as a resource for quantum information processing,
especially in a communication scenario~\cite{teleportation,E91}.

Is there any other quantum-classical boundary,
which would be useful in another scenario?
In many experiments, the quantum system of interest
is held in a local experimental apparatus, such as 
a vacuum chamber and a refrigerator.
In such a situation, can we decide whether or not
the system is quantum in a reasonable sense?

In this paper, we establish a quantum-classical boundary from the viewpoint 
of classical simulatability of a quantum dynamics under decoherence,
which we call a computational quantum-classical (CQC) boundary.
This is motivated by increasing importance of 
computational complexity in physics~\cite{Gefter},
and increasing demands for experimental verification~\cite{Reichardt13}
of complex quantum dynamics, such as 
quantum simulation and quantum annealing~\cite{Dwave11,Dwave14,DwaveNatPhys}.

For this purpose, 
nonlocality or entanglement
is not enough since 
there are a lot of classically simulatable classes of quantum computation,
which can generate highly entangled states
~\cite{Gottesman,matchgates0,BravyiRaussendorf,FujiiMoriIQP}.
Moreover, highly mixed state quantum computation with less entanglement
exhibits nontrivial quantum dynamics~\cite{DQC1,DQC1hardness,DQC12}.
Thus we have to develop a novel criterion,
which determines whether or not the system is classically simulatable. 

Here we consider commuting (diagonal) quantum circuits
preceded and followed by state preparations and 
measurements whose bases are not diagonal.
This setting is quite simple and less powerful than universal 
quantum computation but still exhibits nontrivial quantum dynamics~\cite{IQP0,IQP,FujiiMoriIQP}.
They can be applied, for example, to
a random state generation and a thermalizing algorithm of 
classical Hamiltonian~\cite{NakataDiag}. 
We derive a threshold on the noise strength, below which
the system has quantumness in the sense that
the measurement outcomes 
cannot be simulated efficiently by any classical computer
under some reasonable assumptions.
Hence we call such a region {\it quantum side}.
On the other hand,
if the noise strength lies above another threshold,
the measurement outcomes can be efficiently simulated 
by a classical computer.
We call this region {\it classical side}.
Specifically, when non-constant depth commuting quantum circuits
are followed by single-qubit complete-positive-trace-preserving (CPTP) noises
(or equivalently weighted graph states of a non-constant degree
being subject to single-qubit CPTP noises),
the CQC boundary is given sharply by $q=14.6\%$.
Here $q$ is a noise strength
measured appropriately from the CPTP map
and almost equivalent to the error probability 
on the measurement outcome.f
Even in the case of depth-four circuits,
we show that 
the CQC boundary is sharply
upper and lower bounded 
by $14.6\%$ and $13.4\%$,
respectively.
We also discuss how to verify quantumness in the computational sense
by a single-shot experimental result under some physical
assumptions without relying on any tomographic technique.

In particular, to show intractability of classical simulation in the quantum side,
we utilize the postselection argument introduced by
Bremner, Jozsa and Shepherd~\cite{IQP} and further extend 
it for the system being subject to rather general decoherence.
This extension is crucial for our purpose. 
This is because the original postselection argument holds only for an approximation 
with a multiplicative error.
However, the assumption of the multiplicative error 
or even an additive error with the $l_1$-norm
is easily broken in actual experimental systems,
where noise is introduced inevitably.
If noisy quantum circuits with postselection cannot decide 
post-BQP (or equivalently PP) problems,
hardness of weak sampling with a multiplicative error
would originated from an analog nature of the sampling problems.
If it is true, 
the hardness results 
on sampling would not be physically detectable
like classical analog computing with unlimited-precision real numbers,
which can solve NP complete and even PSPACE complete problems~\cite{AaronsonPhysReal,ClassicalAnalog}.

To tackle this issue, we directly show that
commuting quantum circuits being subject to 
decoherence themselves 
(or MBQC on noisy weighted graph states)
are classically intractable 
if a strength of noise is smaller than a certain constant threshold value.
In doing so, we virtually utilize fault-tolerant quantum computation
to extend the complexity result in an ideal case to 
a noisy case.
To our knowledge, this is the first result
on fault-tolerance of the intermediate classes of quantum computation;
even noisy quantum circuits can decide 
post-BQP (or equivalently PP) complete problems under postselection.
This fact indicates that 
the hardness of the intermediate class consisting of the commuting quantum circuits,
relying on postselection, is 
robust against noise and physically realistic.

On the other hand, classical simulatability in the classical side
is shown by taking a projected-entangled-pair-state (PEPS) picture~\cite{PEPS}.
Not only the separable criteria~\cite{RaussendorfBravyi,BarrettThermal}, 
we also develop a criteria for
the shared entangled pair to become a
convex mixture of stabilizer states.
This allows us to show classical simulatability of 
highly entangling operations.
We explicitly construct a classical algorithm
that simulate noisy commuting quantum circuits,
which would be useful to simulate noisy and complex physical dynamics
with minimum computational effort.

The rest of the paper is organized as follows.
First, 
we preliminarily introduce 
commuting quantum circuits and 
the postselection argument developed on them.
In Sec.~\ref{sec3},
we provide a generic threshold theorem for postselected 
quantum computation,
which shows robustness of the postselected argument 
against decoherence.
In Sec.~\ref{sec4},
we derive a CQC boundary, which sharply 
separates the classically simulatable and not simulatable 
regions. 
In Sec. 4~\ref{sec5},
we provide an experimental verification scheme,
which determines the system is classically simulatable or not, 
based on locality and homogeneity of noise.
In Sec. 5~\ref{sec6},
we generalize the results
into general commuting circuits with arbitrary rotational 
angles to draw a complexity landscape of the system.
Section ~\ref{sec7}
is devoted to discussion.

\section*{Commuting quantum circuits and postselection}
\label{sec2}
\begin{figure}
\centering
\includegraphics[width=85mm]{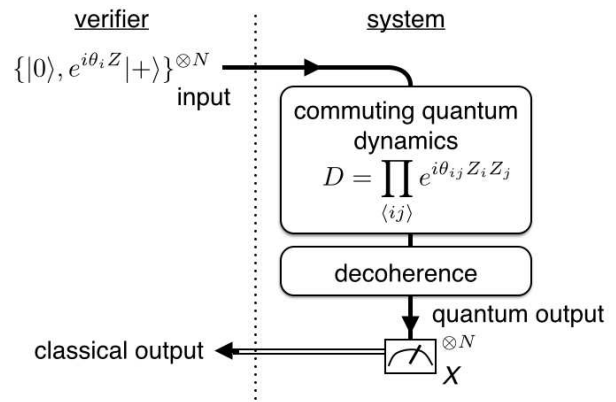}
\caption{Commuting quantum circuits consist
of the input states, commuting gates followed by 
decoherence, and the $X$-basis
measurements. In the verification,
input states 
are under control of the verifier, 
and noisy commuting quantum gates
are verified by using the measurement outcomes.
}\label{fig0}
\end{figure}
The commuting quantum circuit
consists of an input state, 
dynamics, and measurements as shown in Fig.~\ref{fig0}.
The input state is given as a product state
of $N$ qubits, $\{ |0\rangle, e^{i \theta Z} |+\rangle\}^{\otimes N}$,
which are assumed to be arranged on a lattice $\mathcal{L}$.
The dynamics $D$ consists of 
commuting two-qubit gates 
$D= \prod _{\langle ij \rangle} e^{i \theta _{ij} Z_i Z_j}$, 
where $i$th and $j$th qubits are connected on a lattice $\mathcal{L}$,
and $A_i$ indicates an operator $A$ acting on the $i$th qubit.
The measurements are done in the $X$-basis.
By choosing an input state of a qubit to be $|0\rangle$,
the commuting gates acting on the qubit 
can be effectively canceled.
(Or equivalently, instead of using the input $|0\rangle$,
we may change the lattice structure.)
Since $D|+\rangle ^{\otimes N}$ is 
a weighted graph state~\cite{GraphState},
the system can also viewed as 
MBQC on weighted graph states.
In this case, instead of the input $|0\rangle$,
we measure the qubit in the $Z$-basis.
Other qubits are measured on $xy$-plane.
Below, we will mainly expand our argument 
in quantum commuting circuits,
but we can always interpret the results 
in MBQC on the weighted graph states.

The commuting quantum circuits
apparently belong to the class IQP~\cite{IQP0,IQP}.
Since adaptive measurements are not allowed,
the commuting quantum circuits (or IQP) are less 
powerful than universal quantum computation.
However, 
if we are allowed to use postselection,
we can simulate universal MBQC
by choosing the measurement outcomes 
that do not need any feedforward operation.
This implies that 
the postselected commuting quantum circuits 
are as powerful as probabilistic polynomial-time computation (PP)
by virtue of post-BQP=PP theorem~\cite{postBQP}.
As shown in Ref.~\cite{IQP}, 
if the output $\{ m_k \}$
of such a commuting quantum circuit can be efficiently sampled
with a multiplicative error $1<c<\sqrt{2}$ 
using a classical randomized algorithm,
the polynomial hierarchy (PH) collapses at the third level~\cite{IQP}.

The above postselection argument has been quite 
successful, showing classical intractability of 
the experimentally feasible 
intermediate models,
such as commuting quantum circuits (so-called IQP)~\cite{IQP}, liner optics (boson sampling)~\cite{boson}, and highly-mixed state quantum computation (deterministic quantum computation with one-clean qubit~\cite{DQC1})~\cite{DQC1hardness}.
However, the above argument holds only for sampling 
with a multiplicative approximation error,
which is experimentally hard to achieve and verify.
This is the reason why researchers
have also argued the intractability with an additive error
under some plausible complexity conjectures~\cite{boson,IQPadditive}.
However, the hardness is characterized by 
a constant additive error measured by 
$l_1$-norm of the output probability distribution.
This is unsatisfactory in a physically realistic scenario, 
where each gate element
is subject to a noise of a constant strength,
and hence an additive error bound in the sense of $l_1$-norm
is easily broken.

\section{Postselected threshold theorem}
\label{sec3}
Here, we will show that intractability 
of commuting quantum circuits is robust against noise.
Specifically, the hardness is characterized by 
the noise strength measured by 
an appropriate operator norm of the commuting circuits
followed by noise.
To this end, 
we introduce an equivalent reduction; noise in the output probability distribution,
which would spoil the multiplicative approximation,
is regarded as a part of a quantum task 
and an ideal sampling of it is executed.
Then we show that 
such a noisy quantum task itself can solve 
a PP-complete (or equivalently post-BQP-complete) problem.
Importantly, we do not assume any detail of the noise
as long as it is given by spatially-local CPTP map
and criteria is given with respect to a noise strength
measured by a relevant superoperator distance measure.
To prove this, we virtually utilize fault-tolerant quantum computation
as explained below in detail.

The postselected commuting quantum circuits 
can simulate universal measurement-based quantum 
computation (MBQC) as mentioned before.
This implies that 
topologically protected MBQC on a three-dimensional (3D) cluster state
can also be simulated~\cite{RaussendorfAnn,RaussendorfNJP,ReviewFujii}.
The reason why we employ topologically protected MBQC
is that it exhibits high noise tolerance
while the resource state can be generated simply
by a depth-four commuting quantum circuit.
This property is useful in various situations
to show quantum computational capability in the presence of noise~\cite{RaussendorfBravyi,BarrettThermal,FujiiMoriTher,LiKwek,BlindMoriFuji,FujiiNakata}.
Moreover, we can also calculate (a lower bound of) 
the threshold value
rigorously using the self-avoiding walks~\cite{Dennis}.
(As a review of topologically protected MBQC, see Ref.~\cite{ReviewFujii} for example.)
\begin{figure}
\centering
\includegraphics[width=85mm]{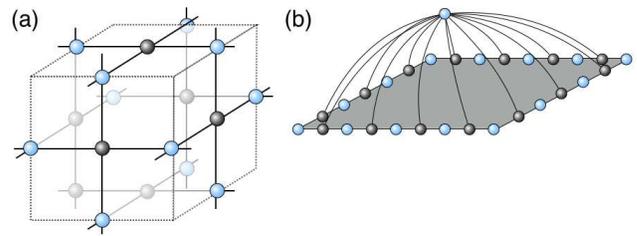}
\caption{
The graphs representing the commuting quantum circuits.
(a) A unit cell of the RHG lattice $\mathcal{L}_{RHG}$ (a graph of degree four) represents 
a depth-four commuting quantum circuit. 
(b) A non-constant depth commuting quantum circuit for a direct magic state injection.
}\label{fig4}
\end{figure}

We consider 
commuting quantum circuits on
a Raussendorf-Harrington-Goyal (RHG) lattice $\mathcal{L}_{\rm RHG}$,
where each face center qubit is connected with 
four surrounding edge qubits on a cubic lattice
as shown in Fig.~\ref{fig4} (a).
This corresponds to a depth-four commuting quantum circuit.
We restrict our attention to two-qubit commuting gates
with $\theta _{ij}=\pi/4$, i.e. a maximally entangling case 
(later we will consider general two-qubit commuting gates).
Then the dynamics $D$ generate 
the cluster state on the RHG lattice.
Specifically, 
input states are chosen to be $|0\rangle$, $|+\rangle$,
and $e^{i (\pi /8) Z}|+\rangle$
to create the defect, vacuum, and singular-qubit regions, respectively.
If the noise level is sufficiently smaller than 
the threshold value for topologically protected MBQC,
classical simulation of such a noisy commuting quantum circuit is also hard.
More importantly, we can go further beyond the standard noise threshold 
by virtue of postselection.
Since we are allowed to use postselection,
we can execute error detection, 
without any cost, which discards 
any possible error events.
Since the noise threshold for error detection 
is much higher than the noise threshold for error correction~\cite{KnillNature,Knilldetection,FYCluster,FYClusterTopo},
intractability of the commuting quantum circuits is much more robust against noise than
the standard universal quantum computation.

We model the noise as a $k$-spatially-local CPTP map
$\mathcal{N}_j$. 
Here $\mathcal{N}_j$ is a super-operator 
acting on the $j$th qubit and its at most $(k-1)$th nearest neighbor qubits
on the RHG lattice $\mathcal{L}_G$.
We are assumed not to know the detail of the noise
except that it is spatially local.
Nevertheless we can show the following theorem.
\begin{theo}[Postselected threshold]
Suppose the dynamics $D$ is followed by 
arbitrary $k$-spatially-local noise $\prod _{j=1}^{N} \mathcal{N}_j$. 
There is a constant threshold $\epsilon _{\rm th} $ 
such that if $\dianorm{\mathcal{N}_j - \mathcal{I}} \leq \epsilon _{\rm th}$,
then efficient classical simulation of
the output of the noisy commuting quantum circuits is impossible 
unless the PH collapses at the third level.
Here $\dianorm{\cdot}$ denotes the diamond norm of the super-operators~\cite{Diamond}.
\label{theo_PostThreshold}
\end{theo}

\noindent{\it Proof:}
The defect regions are introduced by choosing the 
input state to be $|0\rangle$.
The magic state injection can be done by using 
the input state $e^{i (\pi /8) Z}|+\rangle$.
By the $X$-basis measurements,
we can perform topologically protected MBQC.
The postselection is utilized to 
avoid feedforward operations of MBQC.
In the vacuum region,
we obtain a parity $S_u = \bigoplus _{i \in \partial u} \tilde m_i$ 
of six measurement outcomes of the face qubits on a unit cube $u$,
as an error syndrome.
The postselection is further employed
not only to choose the measurement outcomes 
with no feedforward operation
but also to discard the erroneous events
with odd parities,
i.e., $S_u =1$.

Below we will bound the logical error probability 
by modifying the argument developed in Ref.~\cite{Dennis}
under the condition of all even parities, $S_u=0$.
We first decompose the $k$-spatially-local noise $\mathcal{N}_j$ into 
\begin{eqnarray}
\mathcal{N}_j  = (1-\epsilon) \mathcal{I} + \mathcal{E}_{j}  ,
\end{eqnarray}
where $\mathcal{I}$ is an identity super-operator 
and $\epsilon \equiv \max _j  \dianorm{\mathcal{N} - \mathcal{I} }$.
$\mathcal{E}_j$ is a residual $k$-spatially-local super-operator
and may no longer be a CPTP map.
Note that we have $\dianorm{\mathcal{E}_j} \leq 2\epsilon$.
The density matrix is divided
into sparse and faulty part
\begin{eqnarray}
\rho_{\rm noisy} &\equiv&   (1-\epsilon)^N \prod _{j=1}^{N}[\mathcal{I}+\mathcal{E}_j/(1-\epsilon)] \rho
\nonumber \\
&=& (1-\epsilon)^N \sum _{\eta=0}^{N} \left[  \mathcal{I} \sum _{(j_1,...,j_\eta)} \left( \prod _{l=1}^{\eta} \frac{\mathcal{E}_{j_l} }{1-\epsilon}\right) \rho \right]
\nonumber \\
&=& \rho _{\rm sparse}+\rho_{\rm faulty},
\end{eqnarray}
where the summation $\sum _{(j_1,...,j_\eta)}$
is taken over all possible configurations $(j_1, ..., j_\eta)$ 
($j_k =1,...,N$, $j_k \neq j_{k'}$).
The faulty part $\rho _{\rm faulty}$
consists of a super-operator $\prod _{l=1}^{\eta} \mathcal{E}_{j_l}$
whose support $\cup _{l=1}^{\eta} {\rm supp}(\mathcal{E}_{j_l})$ 
covers a logical error.
The operator $\rho _{\rm sparse}$ never 
contributes to the logical error probability
under postselection.
The logical error probability, i.e.,
the $l_1$-distance between the probability distributions for 
the ideal state $\rho _{\rm ideal}$ and 
the noisy state $\rho _{\rm noisy}$
can be bounded by 
the operator-$1$ norm of the faulty operator $\rho _{\rm faulty}$~\cite{AliferisQEC}:
\begin{eqnarray}
&&\sum _{\nu} |P_{\rm ideal}(\nu)- P_{\rm FT}(\nu |{\rm post})|
\nonumber \\
&=&\sum_{\nu} | {\rm Tr} [M_{\nu} (\rho _{\rm ideal} - \tilde \rho_{\rm noisy} /{\rm Tr}
[\tilde \rho_{\rm noisy}])]|
\nonumber \\
&=&
\sum_{\nu} | {\rm Tr} [M_{\nu} (\rho _{\rm ideal} - (\tilde \rho_{\rm sparse}+\tilde \rho _{\rm faulty}) /{\rm Tr}
[\tilde \rho_{\rm noisy}]) ]|
\nonumber \\
&\leq & 2\| \tilde \rho _{\rm faulty} \|_1/(1-\epsilon)^N \leq 2\| \rho _{\rm faulty} \|_1/(1-\epsilon)^N
\label{eq_l1norm}
\end{eqnarray}
where $M_{\nu}$ is the projector for the final measurement,
and $\tilde \rho = P^{\rm post} \rho P^{\rm post}$ 
is an unnormalized postselected 
density matrix with $P^{\rm post}$ being the projection 
to the postselection event.
To obtain the last line, we used the fact that the 
postselection probability is lower bounded as follows:
$ {\rm Tr} [\tilde \rho _{\rm noisy}] \geq (1-\epsilon)^N$.
Below we will show that
Eq. (\ref{eq_l1norm}) is upper bounded by an exponentially decreasing function
by evaluating $\| \rho _{\rm noisy} \|_1$.

To count all configurations $(j_1,...,j_\eta)$ in $\rho _{\rm faulty}$,
which possibly cause logical errors,
below we will assume a super-operator $\mathcal{E}_j$
can put arbitrary errors 
on its support qubits $\in {\rm supp} (\mathcal{E}_{j})$ in the most adversarial way.
$\mathcal{E}_j$ originated from a $k$-spatially local noise $\mathcal{N}_j$
can put at most $(2k-1)$ adversarial Pauli errors around the $j$th qubit.
Moreover, the noise $\prod _{j \in A} \mathcal{E}_j$ with 
a set $A$ can put arbitrary adversarial Pauli errors
on the qubits on $\cup _{j \in A} {\rm supp} (\mathcal{E}_j)$.
This allows us to employ the conventional 
counting argument of the number of self-avoiding walks~\cite{Dennis}.

The faulty part is decomposed into contributions with respect to 
error chains $\mathcal{L}$ of length $L$:
\begin{eqnarray}
\|\rho _{\rm faulty} \|_1\leq \sum _{L=L_d}^{N}\sum _{\mathcal{L}|
|\mathcal{L}|=L} \| \rho _{\rm faulty}^{\mathcal{L}}\|_1,
\end{eqnarray}
where $L_d$ is the minimum size of the defects.
Denoting the set of configurations that 
possibly cause error chains $\mathcal{L}$ 
of length $L$ by 
$I_{\mathcal{L}}\equiv \{ (j_1 ,..., j_\eta) | \mathcal{L} 
\subset  \cup _{l=1} ^{\eta } {\rm supp}(\mathcal{E}_{j_l})\}$,
we have
\begin{eqnarray}
\rho _{\rm faulty}^{\mathcal{L}}= (1-\epsilon)^N 
\sum _{(j_1 ,..., j_\eta) \in I_{\mathcal{L}}} \prod _{l=1}^{\eta } 
\frac{\mathcal{E}_{j_l}}{1-\epsilon}\rho .
\end{eqnarray}
Since $\mathcal{E}_{j_l}$ is $k$-spatially local, $\eta$ have to be at least $r \equiv \lceil L/(2k-1) \rceil$.
Accordingly,
\begin{eqnarray}
&& \| \rho _{\rm faulty}^{\mathcal{L}}\|_1
\nonumber \\
&\leq & (1-\epsilon )^{N}\sum _{\eta = r} ^{L(2k^2-2k+1)} \sum _{(j_1,...,j_\eta)| I_\mathcal{L}} \prod _{l=1}^{\eta} \frac{\dianorm{\mathcal{E}_{j_l}}}{1-\epsilon} 
\nonumber \\
&\leq & (1-\epsilon )^{N} \sum _{\eta=r}^{L(2k^2-2k+1)} 
\left(\begin{array}{c}
L(2k^2-2k+1)
 \\
\eta
\end{array} \right)
\left(\frac{2\epsilon}{1-\epsilon}\right)^{\eta}
\label{eq:fail_prob}
\\
&< & (1-\epsilon)^N\left(\frac{2\epsilon}{1-\epsilon}\right) ^r 2^{L(2k^2-2k+1)} ,
\end{eqnarray}
where we used the properties of the diamond norm~\cite{Diamond}.
The number of error chains 
of length $L$ in the 3D lattice 
can be bounded by
$N (6/5) 5^{L}$ from the number 
of 3D self-avoiding walks.
Thus the logical error probability is bounded by
\begin{eqnarray}
&&\| \rho _{\rm faulty} \|_1 
/(1-\epsilon)^N 
\nonumber \\
&<& N(6/5) \sum _{L=L_{d}}^{N} 
\left[ 5 \cdot 2^{2k^2-2k+1}  \left( \frac{2\epsilon}{1-\epsilon} \right)^{1/(2k-1)} \right]^{L}.
\end{eqnarray}
The total failure probability decreases exponentially in the defect size $L_d$,
if $2\epsilon /(1-\epsilon)< 1/( 5 \cdot 2^{2k^2-2k+1} )^{2k-1}$.
Since $k$ is a finite constant, there is a constant threshold on $\epsilon$,
below which Clifford gates are topologically protected 
under postselection.
Furthermore, if $\epsilon$ is sufficiently smaller than
a certain constant value,
the magic state distillation for universal quantum computation~\cite{Magic,Reichardt05}
can also be done under postselection.
The logical error probability 
of the magic state can be reduced exponentially
with a polynomial overhead.
Accordingly 
there exists a postselected noise threshold $\epsilon_{\rm th}$,
below which we can perform fault-tolerant quantum computation,
i.e., the postselected logical error probability 
decreases exponentially.
That is, for an arbitrary output $\nu$,
we have 
\begin{eqnarray}
| P_{\rm FT}(\nu |{\rm post})
- P_{\rm ideal}(\nu) | < 2^{-\kappa},
\end{eqnarray}
where the overhead $N={\rm poly}(n,\kappa)$
is polynomial both
in the size $n$ and the exponent $\kappa>0$ 
of the logical error probability.

Let us consider an
output of an ideal quantum circuit of size $n$,
$P_{\rm ideal}(x,y) = {\rm Tr} [M_{x,y} \rho _{\rm ideal} ]$,
where $x \in \{ 0,1\}$ and $y \in \{0,1\}$ are
decision and postselection registers, respectively.
Its postselected fault-tolerant version
is $P_{\rm FT}(x,y | {\rm post}) =
{\rm Tr} [M_{x,y} \tilde \rho _{\rm noisy} ]/
{\rm Tr}[\tilde \rho _{\rm noisy}]$.
Now we simulate postselected quantum computation
$P_{\rm ideal}(x|y=0)$ by postselected fault-tolerant 
quantum computation $P_{\rm FT}(x|y =0 , {\rm post})$.
The postselected probability distribution 
is obtained as
\begin{eqnarray}
&& | P_{\rm FT}( x|  y ,{\rm post} ) -
P_{\rm ideal}(x|y)| 
\nonumber \\
&\leq& 
\left| \frac{P_{\rm FT}(x, y | {\rm post}  )}{P_{\rm FT}( y | {\rm post}  )}
- \frac{P_{\rm ideal}(x,y)}{P_{\rm FT}(y | {\rm post}  )} \right|
\nonumber \\
&&+
\left|\frac{P_{\rm ideal}(x,y)}{P_{\rm FT}(y |{\rm post}  )} - 
\frac{P_{\rm ideal}(x,y)}{P_{\rm ideal}(y)}\right|
\nonumber \\
&\leq& 
\frac{1}{P_{\rm FT}(y | {\rm post}  )}
\left|P_{\rm FT}(x, y | {\rm post}  )
- P_{\rm ideal}(x,y) \right|
\nonumber \\
&&+
\left|\frac{1}{P_{\rm FT}(y | {\rm post}  )} - \frac{1}{P_{\rm ideal}(y)}\right|
\nonumber \\  
&\leq &
 \frac{2^{-\kappa}\left(1+2^{6n+4} \right)}{2^{-6n-4}-2^{-\kappa}} \equiv \epsilon(\kappa,n).
\end{eqnarray}
Here we utilized the fact that 
the postselection with an exponentially small probability 
$P_{\rm ideal}(y)>2^{- 6n-4}$ is enough to solve a PP complete problem
of the size $n$ (see Appendix~\ref{AppostBQP=PP} for the detail).
We can always choose $\kappa$ as a polynomial function of $n$
such that $\epsilon(\kappa,n) <1/2$ for an arbitrary $n$. 
Thus postselected noisy commuting quantum circuits 
can solve post-BQP complete (or equivalently PP complete) problems.
This indicates that 
the noisy commuting quantum circuits with postselection
are as hard as PP, and hence no efficient classical simulation exists
unless the PH collapses at the third level.
\hfill $\square$

From the above theorem, we can 
induce the following corollary:
\begin{coro}
Let us consider
noisy commuting (IQP) circuits consisting of 
$|+\rangle$ state preparations,
single-qubit $Z$ rotations and $X$-basis
measurements followed by 
single-qubit CPTP noises, and two-qubit 
commuting gates followed by two-qubit CPTP noises.
There exists a constant threshold value
on the noise strength (the distance with the identity map measured 
by the diamond norm),
below which
classical sampling (with exact or 
with an multiplicative error $1<c<\sqrt{2}$) of 
the noisy commuting circuits 
is hard unless the PH collapses to the third level.
\end{coro}
Note that 
the above CPTP noise of a constant noise strength can easily 
breaks the bounds on the multiplicative or additive error with the $l_1$-norm,
which are employed in the original arguments~\cite{IQP,IQPadditive}.

\noindent{\it Proof:}
Finite depth commuting circuits are enough 
to construct a topologically protected MBQC 
on the 3D cluster state.
Therefore, the single- and two-qubit CPTP noises
can always be written as $k$-spatially-local noises
after the commuting gates. 
Then we can employ Theorem~\ref{theo_PostThreshold}.
\hfill $\square$

Note that in the above proof,
we directly show the noisy commuting quantum circuits 
with postselection include PP or post-BQP,
instead of showing that they are BQP-complete and further postselection 
boosts them into post-BQP.
If the latter is possible, the statement is somewhat trivial.
However, this is not the case. 
Importantly, even if a computational model $A$ 
is BQP-complete, it does not directly lead 
that $A$ with postselection is as powerful as post-BQP.
For example, BQP-complete problems such as
approximations of Jones/Tutte polynomials~\cite{AharonovJones,AharonovJonesHard,AharonovTutte} 
and Ising partition functions~\cite{MatsuoFujii}
are more powerful than IQP~\cite{IQP0,IQP} or DQC1~\cite{DQC1} as decision problems,
but would not become post-BQP complete even with the help of postselection.
(See, for example, Ref.~\cite{VandenNestCommuting} for 
the distinction between decision and sampling problems.)
Moreover, since the probability of postselection
is exponentially small,
the logical error probability has to be 
reduced exponentially.
Fortunately, in fault-tolerant theory,
we can reduce the logical error probability 
exponentially with increasing the overhead polynomially.
These facts allow the postselected noisy quantum circuits to decide 
post-BQP complete problems.

Since the dynamics consists only of two-qubit commuting gates of a constant depth,
noises introduced by the input states, the commuting gates, and the measurements can also be 
regarded as a $k$-spatially-local noise as long as they are also local in space.

\section{A sharp CQC boundary}
\label{sec4}
Next we derive a CQC boundary that sharply
divides the classically simulatable and intractable regions of noisy commuting quantum circuits.
To this end,
we consider the simplest case:
the dynamics is homogeneously subject to a single-qubit CPTP map
\begin{eqnarray}
\mathcal{N} \rho = \sum _i W_i \rho W_i ^{\dag},
\end{eqnarray}
where $W_i = \sum _l c_{il} \sigma _l$
with $\sigma _l$ being the Pauli matrices.
Moreover, non-constant-depth commuting quantum circuits are also employed
for the magic state injection.
The latter requirement is relaxed to constant-depth circuits later.

We are supposed to be blind to the detail of 
the noise in experiments. 
Thus we have to transform the CPTP noise into dephasing 
by using a subprotocol as follows.
In the vacuum and singular-qubit regions,
the input state is chosen to be $ X^{\xi _j} Z^{\bar \nu _j } e^{ i \theta _j Z} |+\rangle _j$,
where $\bar \nu _j \equiv \nu_j \bigoplus _{k \in \partial j} \xi _k$ with $\partial j$ being 
neighbors of the $j$th qubit, and
$\{ \xi _j \}$ and $\{\nu _j \}$ are random binary variables 
with probability 1/2.
The measurement outcomes 
are reinterpreted as $\tilde{m}_i = m_i \oplus \nu _i$.
\begin{figure}
\centering
\includegraphics[width=80mm]{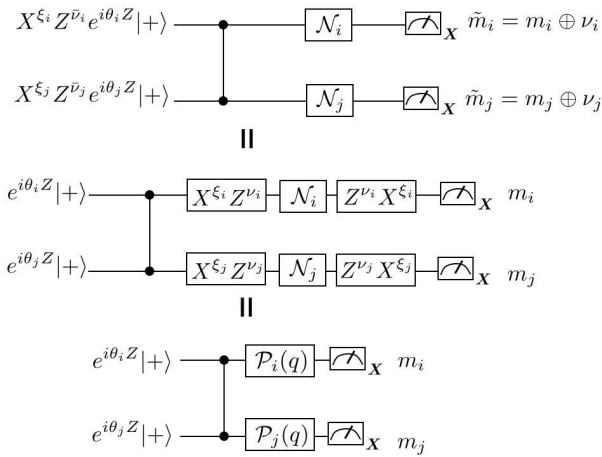}
\caption{
The subprotocol (top)
is equivalent to the circuit where each single-qubit 
CPTP noise is sandwiched by stochastic Pauli operations (middle).
The stochastic Pauli operations depolarize the CPTP noise 
into a stochastic Pauli noise. Since the measurement is done in the $X$-basis,
the stochastic Pauli noise can be given as a dephasing.
}\label{fig3}
\end{figure}
{\red This subprotocol is equivalent to 
the original commuting quantum circuit
where each single-qubit CPTP noise is sandwiched by 
stochastic Pauli operations as shown in Fig.~\ref{fig3}.
These stochastic Pauli operations diagonalize 
the CPTP noise into 
a stochastic Pauli noise~\cite{StandardNoise}.}
Under these operations and using the fact that the measurements 
are done in the $X$-basis,
an arbitrary single-qubit CPTP noise $\mathcal{N}_j$ can be rewritten
as a dephasing~\cite{StandardNoise}:
\begin{eqnarray}
 \mathcal{P}(q) \rho &=& 
(1-q)\rho + q Z\rho Z
\end{eqnarray}
with a dephasing rate 
$
q \equiv \sum _{i,l=2,3} |c_{i l}|^2
$.

In this case $\epsilon = q$ and $\dianorm{E_j} = q$.
From Eq. (\ref{eq:fail_prob}),
the total failure probability is given by
$N (6/5) \sum _{L=d}^{N} [5q/(1-q) ]^{L}$.
Thus the threshold for the topological protection is given by $q= 16.7\%$.
On the other hand, 
if we inject the magic state 
directly to the defect qubit 
by using a non-commuting circuit as shown in Fig.~\ref{fig4} (b),
the error on the injected magic state is 
given solely by the dephasing on the injected qubit.
The threshold for the 
magic state distillation is given by 
$q = (1-\sqrt{2}/2)/2 = 0.146$~\cite{Magic,Reichardt05}.
Thus postselected threshold is given by $14.6\%$.
If $q \leq 14.6\%$,
classical simulation of such a noisy commuting quantum circuit is impossible.
On the other hand,
if $q>14.6\%$,
any input state lies inside the octahedron of 
the Bloch sphere and hence
can be written as a convex mixture of the Pauli-basis states.
{\red The dynamics consists only of Clifford gates. 
The measurements are done in the Pauli-basis.}
Thus the output distribution is classically simulatable
due to the Gottesman-Knill theorem~\cite{Gottesman}.
This indicates that the CQC boundary,
which divides classically simulatable and not simulatable regions, is 
sharply given by $q=14.6\%$ in the present setup.

Next we consider the constant-depth case,
the depth-four commuting quantum circuit shown in Fig.~\ref{fig4} (a).
In this case, we have to take into account 
the noise accumulation on a logical magic state
originated from the low weight errors (see Appendix~\ref{ApMagic} for the detail). 
We count the number of self-avoiding walks causing logical errors
up to the length 14.
The logical $X$ and $Z$ error probabilities 
as functions of $q$ are given by
\begin{eqnarray}
\bar q_X &=& 4q^3+8q^4+52q^5+200q^6+O(q^7),
\label{eq_magic_1}
\\
\bar q_Z &=& q+7q^4+106q^6+O(q^8),
\label{eq_magic_2}
\end{eqnarray}
respectively.
Since the logical $X$ error causes 
an error during magic state distillation 
with probability 1/2,
the threshold for magic state distillation is given by
\begin{eqnarray}
&& \bar q_X /2 + \bar q_Z \leq (1-\sqrt{2}/2)/2
\\ \nonumber
&\Leftrightarrow& q \leq 0.134.
\end{eqnarray}
The higher order contributions of the length 
larger than $14$ is at most $\sim 10^{-5}$ for each,
and hence the threshold almost converges.
Thus if $q<0.134$, postselected fault-tolerant 
quantum computation can simulate post-BQP,
and hence classical simulation of 
the corresponding noisy commuting quantum circuits
is hard.
While there still remains a gap between the classical simulatable region $q>14.6\%$
and the intractable region $q<13.4\%$,
we can obtain a fairly narrow CQC boundary, 
which is valid even for the constant-depth circuits.

Note that in the standard quantum computation,
the threshold for Clifford gates are much lower than
that for the magic state distillation.
Thus the threshold for fault-tolerant universal quantum computation 
is determined by the threshold $~0.0075$ 
for the Clifford gates~\cite{RaussendorfNJP}.
This is also the case in
the earlier work on transitions of quantum computational 
power of thermal states~\cite{BarrettThermal},
where a large gap between
classical and quantum regions exists.
Then, there has been a natural question 
how powerful the system in
the intermediate region is.
Our result provides an answer to this question.
As shown above, 
if we consider the classical simulatability by using the postselection argument,
the threshold, i.e. CQC boundary, is given solely by the distillation threshold of the magic state.
This result is quite reasonable since
the magic state distillation is an essential ingredient for universal quantum computation.

\section{Verification}
\label{sec5}
We have shown that if the noise strength $q$ is smaller than
a threshold value, the corresponding noisy quantum circuits 
cannot be simulated by classical computer unless the PH collapses at the third level.
Thus if we can estimate the rate $q$ in an experiment efficiently
(later we will show how to do this),
the CQC boundary serves as an efficient experimental criterion
that the dynamics has quantumness in a computational sense.
Below, we show how to estimate the dephasing rate $q$
from a single-shot measurement under some physical assumptions.

\begin{theo}[Single-shot verification]
Suppose the noise is given by homogeneous $1$-spatially-local noise.
If the spatial average $\langle S _u \rangle = 1/|S_u| \sum _u S_u$
is larger than 
0.154,
such a noisy commuting quantum circuit is guaranteed to be hard for classical simulation
with a probability exponentially close to 1 in the system size $N$.
\end{theo}

{\it Proof:}
As mentioned previously,
if the $j$th input state is chosen to be $ X^{\xi _j} Z^{\bar \nu _j} e^{ i \theta _j Z} |+\rangle $
randomly, the $1$-spatially-local noise $\mathcal{N}_j$ can be rewritten as 
a dephasing $\mathcal{P}_j(q)$ with the probability $q \equiv \sum _{i,l=2,3} |c_{i l}|^2$.
The parities $\{S_u=\pm 1\}$
are independent binary variables with probability
$[1+S_u(1-2q)^6]/2$.
The spatial average of $S_u$ is calculated to be 
\begin{eqnarray}
\langle S_u \rangle = (1-2q)^6.
\end{eqnarray}
If $q=0.134$, this reads $0.154$.
By virtue of Hoeffding-Chernoff inequality,
if we obtain $\langle S_u \rangle > 0.154$ experimentally,
the probability that $q > 0.154$ is 
exponentially small, and hence 
classical intractability is guaranteed with 
a probability exponentially close to 1. 
\hfill $\square$

The above arguments can be straightforwardly 
generalized into $k$-spatially-local CPTP noises,
if one assumes spatial homogeneity.
As a practice,
let us consider a more realistic noise model,
where the state preparation and measurements 
are followed by a single-qubit depolarizing noise
\begin{eqnarray}
\mathcal{N}^{(1)}= (1-p_1)[I]+ \sum _{A=X,Y,Z} (p_1/3) [A],
\end{eqnarray}
and two-qubit commuting gate is followed by
two-qubit depolarizing noise
\begin{eqnarray}
\mathcal{N}^{(2)}= (1-p_2)[I]+ 
\sum _{A=\{ I,X,Y,Z\}^{\otimes 2} \backslash I} (p_2/15)[A].
\end{eqnarray}
Here $[A]$ indicates a superoperator $A(\cdots)A^{\dag}$.
In this case, the noise operator 
after the depth-four commuting gate is at most 2-spatially-local.
The correlated errors introduced
on each pair of qubits on opposite edges on each face.
The independent and correlated 
error probabilities $q_{\rm ind}$ and $q_{\rm cor}$ can be obtained 
from a straightforward calculation~\cite{RaussendorfAnn}:
\begin{eqnarray}
q_{\rm ind} &=& \frac{1}{2}[1- (1-16p_2/15)^4(1-4p_1/3)^2],
\\
q_{\rm cor} &=& \frac{1}{2}(1-\sqrt{1-16p_2/15}).
\end{eqnarray}
The correlated error
is located between two unit cubes,
and hence errors are independent for each qubit on a unit cell.
Therefore $\langle S_u\rangle$ can be 
given simply by 
\begin{eqnarray} 
\langle S_u \rangle = [(1-2q_{\rm ind}) (1-2q_{\rm cor})^4]^6.
\end{eqnarray}

On the other hand, the threshold on the magic state distillation
has to be modified appropriately by taking correlated noise into account.
For the errors on the singular qubit,
we counted, up to the leading order, the probability $p_s$ of the errors,
which are located solely on the singular qubit
or the weight-four primal chain and hence 
cannot by postselected.
This amounts to be 
$p_s=(8p_2/15+3p_1/3)+ (4p_2/15+2p_1/3)/2$.
For the chains of weight three or higher,
we replace $q$ with $q_{\rm ind}+4q_{\rm cor}+ \sqrt{q_{\rm cor}}$ 
in Eqs (\ref{eq_magic_1}) and (\ref{eq_magic_2}).
This automatically takes 
the weight-two correlated errors;
for example $q^2 = (q_{\rm ind}+4q_{\rm cor})^2
+ 2 (q_{\rm ind}+4q_{\rm cor}) q_{\rm cor} ^{1/2}
+ q_{\rm cor}$,
where the odd order terms of $\sqrt{q_{\rm cor}}$
are unphysical but only worse the threshold.
Note that this substantially overestimates the 
error probability, since some of them can be 
detected and postselected on the dual lattice.
For simplicity, if we take $p_1 = p_2$,
the threshold is given by $p_1=p_2 = 0.0270$,
which corresponds to $\langle S_u \rangle=0.225$.
Note that the postselected threshold $0.0270$ is still higher
than the standard threshold $\sim 0.0075$~\cite{RaussendorfAnn}
for universal quantum computation. 
On the other hand,
if $p_s > (1-\sqrt{2}/2)/2$,
then the noisy magic state becomes 
a convex mixture of the Pauli basis states.
This indicates that 
if $p_1=p_2 > 0.0998$ for the depolarizing noise model, 
the noisy commuting circuits become classically simulatable.
The gap between $0.0270$ and $0.0998$ is originated 
from that the probability $q_{\rm ind}+4q_{\rm cor}+ \sqrt{q_{\rm cor}}$ 
includes the errors that can be postselected
using the correlation between the primal and dual lattices.
Therefore the true threshold for classical intractability 
would be much higher than $0.0270$.

\section{CQC boundary for general commuting circuits}
\label{sec6}
In the previous argument, 
we explicitly utilized the fact that 
the dynamics consists only of C$Z$ gates.
Here we generalize the dynamics 
to two-qubit nearest-neighbor commuting gates
\begin{eqnarray}
D= \prod _{\langle ij\rangle} e^{i \theta _{ij} Z_i Z_j},
\end{eqnarray}
where $\theta _{ij} \in [0,\pi/4]$, and $\prod _{\langle ij \rangle}$ is taken over all nearest-neighbor two qubits.
For simplicity, we assume 
that noise is intrinsically provided as a dephasing $\mathcal{P}(q)$
consider the depth-four commuting quantum circuits.
The lower bound, i.e. classical intractability, with $\theta _{ij}=\pi/4$
is $q=13.4\%$
for the depth-four circuit
($q=14.6\%$ for the higher depth circuit),
since the previous case is a special case of the 
present one.

\subsection{Classical simulatability: PEPS approach}
\begin{figure}
\centering
\includegraphics[width=85mm]{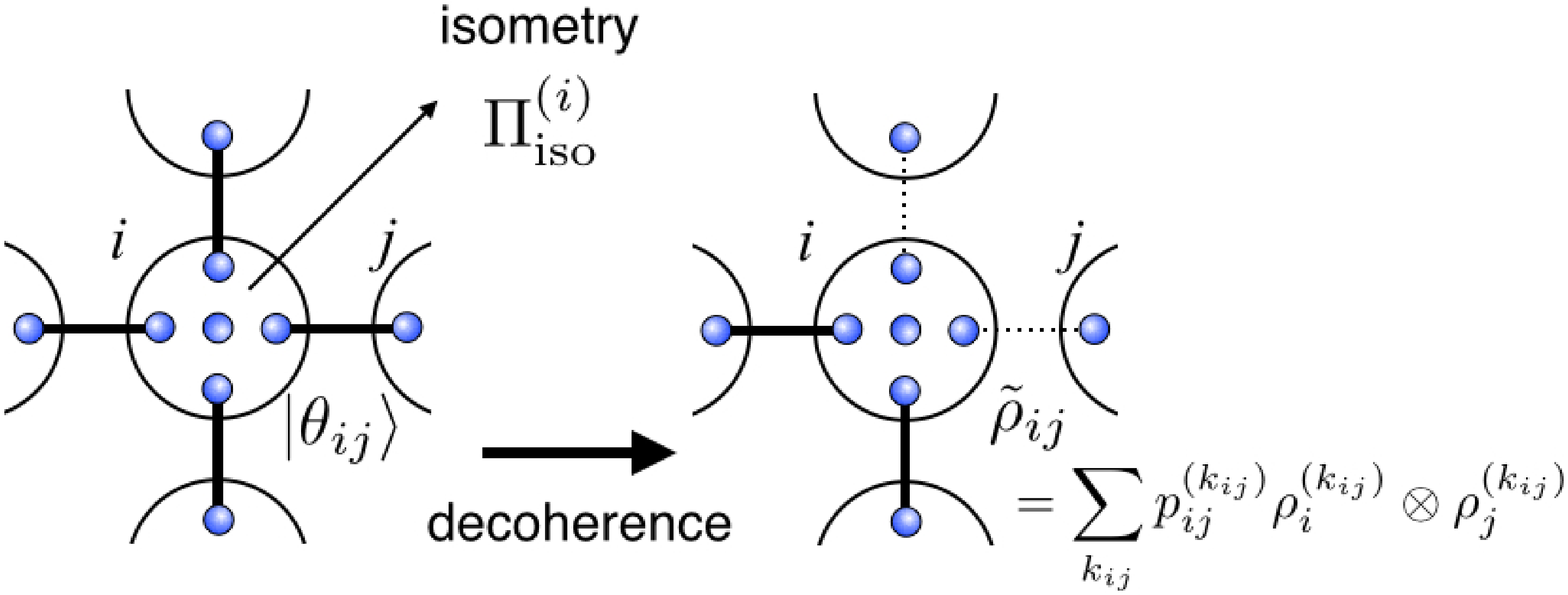}
\caption{A PEPS picture of a depth-four commuting quantum circuit.
Each site denoted by the large circle indicates an original input qubit
of the commuting circuit. An entangled pair shared between the nearest neighbor sites 
is denoted 
by small circles connected by a solid line. The initial input state
is represented as a qubit located at the center of each site.
The dephasing after the commuting gate 
corresponds to disentangling the shared entangled state.
}
\label{fig2}
\end{figure}

Below we will first derive an upper bound of the CQC boundary
showing classically simulatability of 
an arbitrary depth-four commuting quantum circuit under decoherence.
We regard the state before the measurement, which we call a quantum output hereafter, as a PEPS~\cite{PEPS,RaussendorfBravyi,BarrettThermal}.
At the center of the site,
the input state $|\alpha _j\rangle$ is located
to represent an initially rotated single-qubit state.
An entangled pair
\begin{eqnarray}
|\theta _{ij} \rangle \equiv e^{ i \theta _{ij} Z\otimes Z} |+\rangle |+\rangle 
\end{eqnarray}
is shared between nearest-neighbor sites as shown in Fig.~\ref{fig2} (a),
which corresponds to a two-qubit commuting gate.
The isometry (projection) 
\begin{eqnarray}
\Pi _{\rm iso}^{(i)} =  |0\rangle (\langle 0|)^{\otimes 4} +|1\rangle  (\langle 1|)^{\otimes 4},
\end{eqnarray}
defined on each site $i$ reproduces the quantum output
as follows:
\begin{eqnarray}
|\Psi _{\rm out}\rangle \equiv D \bigotimes _{k} |\alpha _k \rangle
= \mathcal{C}
\left(\prod _{k} \Pi _{\rm iso}^{(k)} \right) \bigotimes _{\langle ij\rangle}|\theta _{ij} \rangle \bigotimes _k |\alpha _k \rangle ,
\end{eqnarray}
where $\mathcal{C}$ is a normalization factor.
By denoting $\rho _{\rm out} \equiv |\Psi _{\rm out}\rangle\langle \Psi _{\rm out}|$ and $\rho _{ij} =|\theta _{ij}\rangle \langle \theta _{ij}|$,
the dephasing can be taken as
\begin{eqnarray}
\prod _{i} \mathcal{P}_i (q) \rho _{\rm out} 
&=&  \mathcal{C}^2 \prod _{i}\Pi _{\rm iso}^{(i)}  \Biggl[\left( \bigotimes _{\langle ij \rangle} \mathcal{P}_j (q_{j,i}) \mathcal{P}_i  (q_{i,j}) \rho _{ij} \right) 
\\
&&\otimes
\left(\bigotimes _{k} \mathcal{P}(q_k)|\alpha _k \rangle \langle \alpha _k | \right) \Biggr]
\left(\prod _{i} \Pi _{\rm iso}^{(i)}\right)^{\dag},
\end{eqnarray}
where $q_{i,j}$ and $q_k$
are chosen such that
\begin{eqnarray}
1-2q=(1-2q_k)\prod _{j \in \delta i} (1-2q_{i,j}).
\end{eqnarray}

By choosing $q_{i,j}=q_{j,i}=q^{(i,j)}$,
the dephased entangled pair $\tilde \rho _{ij}$ can be written as
\begin{eqnarray}
\tilde \rho _{ij} &=&\mathcal{P}_i(q^{(i,j)})\mathcal{P}_j(q^{(i,j)})
   \rho _{ij}
   \nonumber \\
&=& \frac{1}{4} \bigl[
II + (1-2q^{(i,j)})\cos 2 \theta _{ij} (IX+XI) 
\nonumber \\
&&
- 
(1-2q^{(i,j)})\sin 2 \theta _{ij}   (ZY+YZ)   + (1-2q^{(i,j)})^2 XX \bigr].
\nonumber \\
\end{eqnarray}
The separability criterion, so-called concurrence, 
for two-qubit mixed state~\cite{Concurrence}
provides the condition
\begin{eqnarray}
(1-2q^{(i,j)}) \leq - \sin 2\theta _{ij} +\sqrt{\sin ^2 2 \theta _{ij} +1}.
\end{eqnarray}
Each site has four nearest-neighbor bonds
since we are considering a depth-four commuting quantum circuits.
If at least two nearest-neighbor bonds per site are made separable for 
as shown in Fig.~\ref{fig2}, the corresponding PEPS can be decoupled into
quasi one-dimensional entangled states (more precisely matrix product states).

After the sampling (see Appendix~\ref{ApSampling} for the detail),
the probability distributions on the 
quasi one-dimensional entangled states
can be evaluated via the matrix products.
Hence the measurement outcomes can be simulated efficiently if 
\begin{eqnarray}
1-2q &\leq&  \left(- \sin 2\theta _{\rm m} +\sqrt{\sin ^2 2 \theta _{\rm m} +1}\right)^2,
\end{eqnarray}
where $\theta _{\rm m} = \max \{ \theta _{ij} \}$ and $q_k=0$ is taken.

\subsection{Classical simulatability: stabilizer mixture approach}
The above argument using the separability 
criteria cannot reproduce classical simulatability 
with $\theta _{ij}=\pi/4$,
where the quantum output is highly entangled.
Next we derive another bound with respect to the Gottesman-Knill theorem.
If
\begin{eqnarray}
(1-2q^{(i,j)}) \leq 
\cos 2 \theta _{ij} + \sin 2 \theta _{ij}
- \sqrt{2 \cos 2 \theta _{ij} \sin 2 \theta _{ij}}
\end{eqnarray}
the entangled pair becomes a convex mixture
of the stabilizer states.
The input state $e^{ i \theta _k Z}|+\rangle$
becomes a convex mixture of Pauli-basis states,
if $1-2q_k \leq  1/(\sin 2 \theta _k + \cos 2\theta _k ) \geq 1/\sqrt{2}$.
Thus if 
\begin{eqnarray}
1-2q &\leq&  \frac{1}{\sqrt{2}}\left(\cos 2 \theta' _{\rm m} + \sin 2 \theta ' _{m}
- \sqrt{2 \cos 2 \theta ' _{\rm m} \sin 2 \theta ' _{\rm m}}\right)^4,
\end{eqnarray}
with $\theta ' _{\rm m} \equiv \max\{ |\theta _{ij} - \pi/4|\}$,
the quantum output becomes a convex mixture of stabilizer states,
on which the Pauli-basis measurements are efficiently classically simulatable.
{\red More precisely,
for each bond, we first choose 
a pure stabilizer state from the convex mixture 
according to the posterior probability conditioned on 
the successful projections as mentioned previously.
In this case, one of the sampled state is given as 
an entangled state
\begin{eqnarray}
\frac{II - (ZY +YZ)+XX}{4}.
\end{eqnarray}
This state can be made separable by using 
the commuting gate $e^{-i (\pi/4)ZZ}$,
which commutes with the isometry.
Thus even in this case, the joint probability of 
successful projections on all sites can be divided into 
probabilities of successful projections on each site.
Then, the sampling with the posterior probability
can be done appropriately.

The $X$-basis measurement of the $i$th qubit
after the isometry (projection) is
equivalent to the measurement of an operator $\prod _{a}X^{(i)}_a$
at site $i$ before the isometry.
Thus the probability distribution of the output of 
the commuting circuits is given by
the probability distribution for $\prod _{a}X^{(i)}_a$
conditioned on obtaining $+1$ eigenvalues 
for all parity operators $\{ Z_a^{(i)}Z_b^{(i)} \}$. 
Such a probability can be evaluated efficiently 
by virtue of the Gottesman-Knill theorem.}

\begin{figure}
\centering
\includegraphics[width=85mm]{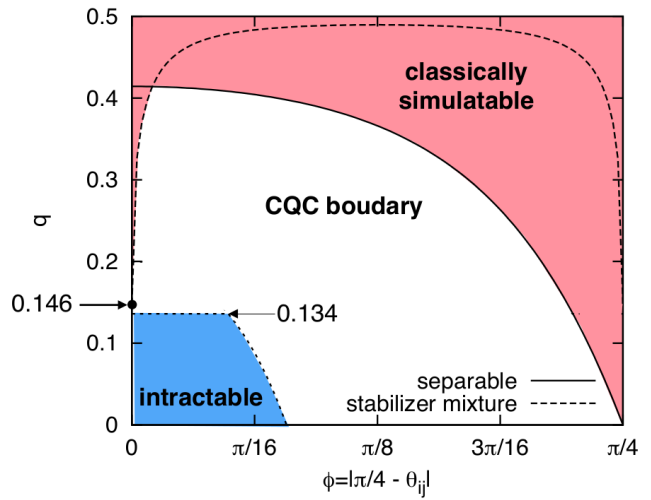}
\caption{A complexity landscape of the depth-four noisy commuting quantum circuits
or MBQC on a weighted graph state of degree four.
Classically simulatable and intractable regions (colored by red and blue respectively)
are shown  
with respect to 
the dephasing strength $q$
and
the rotational angle $\phi =|\pi/4 -\theta _{ij}|$ of 
the two-qubit commuting gates.
The solid line indicates the condition 
for the entangled pair to be a convex mixture of the stabilizer states.
The dashed line indicates the separable criterion
such that the residual entangled pairs can be treated as matrix product states.
Inside the region colored red, the measurement outcomes can be 
classically simulatable efficiently. Inside the region colored blue,
universal fault-tolerant quantum computation 
can be executed under postselection, which implies that 
classical simulation of it is hard.
For the maximally entangling commuting gate with $\phi =0$,
the boundary is sharply given by $0.134$-$0.146$.
}
\label{fig1}
\end{figure}

For simplicity, let us assume 
$\phi =| \pi/4 -\theta _{ij}|$ for all $(i,j)$,
that is, all commuting gates have 
the same entangling power.
Then the separable and stabilizer-mixture criteria
are shown in Fig.~\ref{fig1}.
When $\phi=0.0144$, the dephasing rate $q$ required 
for classical simulation becomes the highest.
In the region $\phi > 0.0144$, the state before the 
measurements is highly entangled but can be written as a convex mixture of 
stabilizer states, and hence
the measurement outcomes can be efficiently simulated. 

\subsection{Classical intractability for general $\theta _{ij}$}
Finally we discuss classical intractability, i.e., lower bound of
the CQC boundary for the general two-qubit commuting gates
with $\phi = |\pi/4 - \theta _{ij}|$ ($\theta _{ij} \in [-\pi/4,\pi/4]$).
The heart of this parameterization is 
that the two-qubit commuting gates are characterized by 
its entangling power; they generate maximally entangled state 
with $\phi =0$ and no-entanglement with $\phi = \pi/4$.
Note that
two different types of two-qubit commuting gates ($\theta_{ij} = \pi/4  \pm \phi$) 
of the same entangling power can be freely chosen.
The choice of the commuting gates is 
inevitable to take the over or under rotation $\phi$
with respect to $\pi/4$ as imperfections as follows.
By choosing $\theta _{ij} = \pi/4 \pm \phi$ randomly
with probability 1/2,
the two-qubit commuting gate 
can be rewritten as 
$e^{ i (\pi/4) Z_i Z_j}$ (equivalent to CZ up to a single-qubit rotation)
followed by a collective dephasing with probability $q(\phi) \equiv \sin^2 \phi$:
\begin{eqnarray}
\rho \rightarrow [1- q(\phi)]  \rho +  q(\phi) ZZ  \rho ZZ.
\end{eqnarray}

Topological quantum error corrections are done 
independently on the primal and dual lattices, respectively.
Suppose the primal lattice is utilized to inject magic states
and perform universal quantum computation and the dual lattice is utilized 
to detect errors.
If a total of the dephasing rates $q$ and $q(\phi)$
is below the topological threshold 20\% (although this is far from tight), that is,
\begin{eqnarray}
 [1-2q(\phi)]^4 (1-2q) \geq 0.6,
\end{eqnarray} 
then the correlated errors are detected and removed 
on the primal lattice. 
Besides, if $(1-2q)< 1/\sqrt{2}$,
magic state distillation succeeds
and hence 
the commuting quantum circuits
can simulate universal quantum computation under postselection.
The classically intractable region $(q,\phi)$, in which 
the dynamics cannot be simulated efficiently unless 
the PH collapses at the third level, is shown in Fig.~\ref{fig1}.

Note that while we here randomly choose the angle $\theta _{ij}= \pi/4 \pm \phi$
to depolarize a commuting gate into a correlated dephsing,
we can also calculate the intractable region for $\theta _{ij}=\pi/4 -\phi$
by taking $e^{-\phi ZZ}$ as a noise and evaluating its diamond norm.

\section{Discussion}
\label{sec7}
Here we have established 
the CQC boundary 
for the commuting quantum circuits under decoherence.
The condition for the system to be 
a convex mixture of the stabilizer states is far from tight
and should be further improved. 
Such a technique required to show classical simulatability 
will be useful to describe a complex and noisy quantum system efficiently. 

On the other hand,
the technique to show classical intractability 
is useful to certify quantumness in an experimentally feasible setup.
It will be interesting to study 
a relation with unconditionally verifiable blind quantum computation~\cite{FitzVeri},
where the quantum tasks are verified without any assumption
but unfortunately have no error tolerance,
meaning that any small error is detected as an evil attack by the quantum server.

The commuting quantum circuits,
which we adopted as an experimentally feasible setup, 
can be readily applicable for a wide range of non-commuting quantum dynamics
by using the Trotter-Suzuki expansion and a path integral method.
It would be interesting to investigate 
the relationship between the present CQC boundary and 
contextuality~\cite{HowardContextuality}, a nonlocal property of quantum systems, 
which has been shown to be relevant for universal quantum computation
via magic state distillation, recently.

While we here addressed fault-tolerance of an intermediate model
of quantum computation
only for commuting circuits,
application of the postselected threshold theorem to 
another intermediate models such as boson sampling and DQC1
might be possible~\cite{boson,DQC1,DQC1hardness,DQC12}.
Specifically, 
there are fault-tolerant models of linear optical quantum computation
~\cite{KLM,DawsonNielsen,FTProb,LiBenjamin},
we could, in principle, apply the postselected threshold theorem
for linear optical quantum computation.
It would be interesting to see 
how it behaves against various sources of noise such as a multi-photon effect and photon loss~\cite{Rohde2014}.

\acknowledgements 
KF is supported by JSPS Grant-in-Aid for Research Activity Start-up 25887034.
ST is supported by the Funding Program for World-Leading Innovative R\&D on Science and Technology (FIRST Program).

\appendix

\section{Exponentially small logical error probability 
is enough to solve postBQP=PP}
\label{AppostBQP=PP}
Here we briefly review post-BQP = PP theorem by Aaronson~\cite{postBQP}
and show that postselection with at most exponentially small
probability is enough to solve a PP-complete problem.
Let $f:\{0,1\}^n \rightarrow \{ 0,1\}$ be an 
efficiently computable Boolean function
and $s=|\{ x: f(x)=1\}|$.
To show PP-completeness,
it is enough to decide whether $s < 2^{n-1}$ or $s\geq 2^{n-1}$.
To this end, we first prepare $2^{-n/2} \sum _{x\in \{0,1\}^n}  
|x\rangle |f(x)\rangle$.
After the Hadamard transformations, 
the first $n$ qubits are measured in the $Z$ basis, and 
we obtain $x=0...0$ with probability at least $1/4$.
The post-measurement state ($|0\rangle^{\otimes n}$ is omitted hereafter)
\begin{eqnarray}
|\psi \rangle = 
\frac{(2^n -s )|0\rangle + s|1\rangle}{\sqrt{(2^n-s)^2+s^2}}
\end{eqnarray}
is entangled with another ancilla qubit 
$\alpha |0\rangle +\beta |1\rangle$ ($|\alpha|^2 + |\beta |^2=1$) as
\begin{eqnarray}
\alpha |0\rangle |\psi\rangle + \beta |1\rangle H |\psi \rangle,
\end{eqnarray}
where $\beta /\alpha = 2^{k}$ with $k \in [-n,n]$ being an integer.
Then postselection on the second qubit by $|1\rangle$
yields
\begin{eqnarray}
|\phi _k \rangle =  
\frac{s\alpha |0\rangle + \beta (2^n -2s)/\sqrt{2}|1\rangle}{\sqrt{(2^n-s)^2 + s^2}}.
\end{eqnarray}
Then if $2^n -2s \leq 0$, i.e.,  $s \geq 2^{n-1}$,
the state never lies in the first quadrant. 
Otherwise, $|\phi _k \rangle$ can be made close to $|+\rangle$
by an appropriate $k$. This separation can be enough to 
then we can decide whether $s < 2^{n-1}$ or $2^{n-1} \leq s$
(see Ref.~\cite{postBQP} for the detail).

The probability of the above postselection 
is calculated to be 
\begin{eqnarray}
  \frac{s^2 + 2^{2k-1} (2^n -2s)^2}{(1+2^{2k})[s^2+(2^n -s)^2]} 
  &>& \frac{1}{2^{2n}(2^{2n}+1)(2+2^{2n+2})} 
 \nonumber \\
  &>& 2^{-6n-4},
\end{eqnarray}
where we used that $2^{-n} \leq 2^k \leq 2^{n}$ and $0 \leq  s \leq 2^{n}$. 
Thus postselection with an exponentially small probability 
$2^{-6n-4}$ is enough to decide a PP-complete problem
of the size $n$.
Let us define postBQP${}^{*}$ as a restricted postselected quantum computation class
whose probability for postselection is lower bounded by $2^{-6n-4}$
in the size $n$ of the problem.
Now we have postBQP${}^{*}$=PP.

Let $P_{\omega} (x,y_1)$ is the output probability distribution 
of $C_\omega$
for uniformly generated quantum circuits $\{C_\omega\}$,
where $x$ and $y_1$ are decision and postselection ports,
respectively.
Let $P(x,y_1,y_2)$ is the output probability distribution
of an element of uniformly generated noisy quantum circuits (possibly
followed by polynomial-time classical computation
to decode the logical information),
where $x$ and $y_{1,2}$ are decision and two postselection ports,
respectively.
Then we can show the following lemma:
\begin{lemm}
For any quantum circuit $C_\omega$,
if there exists a noisy quantum circuit of 
the size $N={\rm poly}(n,\kappa)$
with $n$ being the size of $C_w$
such that
\begin{eqnarray}
|P(x,y_1|y_2=0) - P_w(x,y_1)| < e^{-\kappa},
\end{eqnarray}
then weak classical simulation
with the multiplicative error $\epsilon <\sqrt{2}$ of 
such a uniform family of the noisy quantum circuits
is impossible unless the PH collapses to 
the third level.
\end{lemm}
Here weak classical simulation 
with a multiplicative error $\epsilon$
of the noisy quantum circuits means that 
the classical sampling of $\{m_k\}$
according to the probability distribution 
$P^{\rm ap}(\{m_k\})$ 
that satisfies
\begin{eqnarray}
 (1/\epsilon) P(\{m_k\})< P^{\rm ap}(\{m_k\}) < \epsilon P(\{m_k\}),
\end{eqnarray}
where $P(\{m_k\})$ is 
the output probability distribution 
of the noisy quantum circuit.
\\
{\it Proof:}
A language $L$ is in the class
postBQP${}^{*}$ iff there 
exists a uniform family 
of postselected quantum circuits $\{C_\omega \}$
with a decision port $x$ and postselection port $y_1$
such that
$P_{\omega}(y_1=0) > 2^{-6n-4}$, and
\begin{eqnarray}
\textrm{if } \omega \in L, 
P_\omega (x |  y_1=0) \geq 1/2 +\delta 
\\
\textrm{if } \omega \notin L, 
P_\omega (x |  y_1=0) \leq 1/2 -\delta ,
\end{eqnarray}
where $\delta$ can be chosen arbitrary such that $0 <\delta <1/2$.
Now we have
\begin{eqnarray}
&&\left| 
P(x|y_1 =0,y_2=0)-P_w(x|y_1=0)
\right|
\nonumber \\
&< &
\left|
P(x,y_1|y_2=0)\left(
\frac{1}{P(y_1=0|y_2=0)}
-
\frac{1}{P_{\omega}(y_1=0)}
\right)
\right|
\nonumber \\
&&+
\left|
\frac{P(x,y_1|y_2=0)-P_{\omega}(x,y_1)}{P_w(y_1=0)}
\right|
\nonumber  \\
&<&
\frac{2e^{-\kappa}}{P(y_1=0|y_2=0)P_{\omega}(y_1=0)}
+
\frac{e^{-\kappa}}{P_{\omega}(y_1=0)}
\nonumber \\
&<&
\frac{2e^{-\kappa}}{(P_{\omega}(y_1=0)-e^{-\kappa})P_{\omega}(y_1=0)}
+
\frac{e^{-\kappa}}{P_{\omega}(y_1=0)}.
\end{eqnarray}
Since $P_{\omega}(y_1=0)>2^{-6n-4}$,
we can choose $\kappa = {\rm poly}(n)$
such that  
$\left| 
P(x|y_1 =0,y_2=0)-P_{\omega}(x|y_1=0)
\right| <1/2$.
The resultant size of the noisy quantum 
circuit is still polynomial in $n$.
From the definition (robustness against the bounded error) 
of the class postBQP${}^{*}$ (as same as postBQP),
the postselected noisy quantum circuit 
can decide problems in postBQP${}^{*}$=PP
(recall that we can freely choose $0<\delta <1/2$).
Thus postselected quantum computation 
of such noisy quantum circuits is 
as hard as PP,
and hence cannot be weakly simulated with 
the multiplicative error $\epsilon <\sqrt{2}$
unless the PH collapses to the third level.
\\
\hfill $\square$

\section{Sampling method}
\label{ApSampling}
In a classical simulation,
we have to take into account the success probability of the
projections for the PEPS.
Suppose the dephased entangled pair 
is decomposed into separable states as follows:
\begin{eqnarray}
\tilde \rho _{ij} = \sum _{k} p_{ij}^{(k_{ij})} \rho ^{(k_{ij})}_{i}
\otimes \rho^{(k_{ij})} _{j}.
\end{eqnarray}
To handle the success probability of projections,
we have to sample separable states 
$\{ \rho^{(k_{ij})}_{ij} \equiv \rho ^{(k_{ij})}_{i}
\otimes \rho^{(k_{ij})} _{j} \}_{\rm sep}$ with a posterior probability
conditioned on the success of projections $P_{\rm iso}^{(l)}= |00...0\rangle \langle 
00...0| + |11...1\rangle \langle 11...1|$ on all site $l$:
\begin{eqnarray}
p(\{ k_{ij} \}_{\rm sep}) = 
\frac{
\Tr \left[
\left( \prod _{l} P _{\rm iso}^{(l)} \right) \displaystyle\prod _{\langle S \rangle _{\rm sep}} 
\rho^{(k_{ij})}_{ij} \otimes \rho_{\rm r} \right] \displaystyle\prod _{\langle ij \rangle _{\rm sep}}  p_{ij}^{(k_{ij})}
}
{
\Tr \left[
\left( \prod _{l} P _{\rm iso}^{(l)} \right) \displaystyle\prod _{\langle ij \rangle _{\rm sep}} 
\tilde \rho_{ij}  \otimes \rho_{\rm r}\right],
}
\end{eqnarray}
where $\{\cdot \}_{\rm sep}$ and $\langle \cdot \rangle_{\rm sep}$ 
are sets with respect to the separable bonds,
and $\rho _{\rm r}$ indicates the remaining entangling bonds and 
central qubits $\bigotimes _j |\alpha _j\rangle$ for the input state.
To this end, a separable state $\rho _{ij}^{(k_{ij})}$ 
is sampled independently for each dephased entangled pair $\tilde \rho_{ij}$
according to a posterior probability 
given that the projections at site $i$ and $j$ succeed:
\begin{widetext}
\begin{eqnarray}
\tilde{p}_{ij}^{(k_{ij})}  &=& 
\frac{\Tr \left[P _{\rm iso}^{(i)} 
P _{\rm iso}^{(j)} \left(  \rho _{ij} ^{(k_{ij})} \displaystyle \bigotimes _{j' = \partial i \backslash j} 
 \psi _{i}^{(j')} 
\displaystyle\bigotimes _{i' = \partial j \backslash i} \psi _{j}^{(i')} \otimes |\alpha _i \rangle \langle \alpha _i | \otimes |\alpha _j \rangle \langle \alpha _j | \right)\right] p_{ij}^{(k_{ij})} 
}
{
\Tr \left[P _{\rm iso}^{(i)} 
P _{\rm iso}^{(j)} \left( \tilde \rho _{ij} \displaystyle\bigotimes _{j' = \partial i \backslash j} 
 \psi _{i}^{(j')} 
\displaystyle\bigotimes _{i' = \partial j \backslash i} \psi _{j}^{(i')}\otimes |\alpha _i \rangle \langle \alpha _i | \otimes |\alpha _j \rangle \langle \alpha _j | \right) \right]	
}.
\end{eqnarray}
\end{widetext}
Here if the sampling on bond $(i,j')$ is not yet completed,
$\psi _{i}^{(j')} = {\rm Tr}_{j'} [ \tilde \rho _{ij'}]$
with $\Tr_a[\cdot]$ being a partial trace with respect to qubit $a$.
Otherwise, $\psi _{i}^{(j')}=\rho _i^{(k_{ij'})}$ according to
the sampling result.
Similarly
$\psi _{j}^{(i')} = {\rm Tr}_{i'} [ \tilde \rho _{i'j}]$
or $\psi _{j}^{(i')} =\rho _j ^{(k_{i'j})}$
depending on whether or not the sampling on bond $(i',j)$
is completed.
In other words, the calculation of the posterior probability 
is done with updating the states on the bonds depending on the sampling results.
Since both commuting gate and dephasing operations
commute with the isometry,
the joint probability distribution for the successful projections on all sites
are divided into a product of probabilities of successful projections on each site.
This is also the case for the sampled states, 
since they are separable.
By using these facts, as proved in Ref.~\cite{BarrettThermal},
the sampling according to $\prod _{\langle ij\rangle }\tilde p_{ij}^{(k_{ij})}$ reproduces 
the distribution $p(\{ k_{ij} \})$.

\section{Low-weight error accumulation}
\label{ApMagic}
\begin{figure}
\centering
\includegraphics[width=55mm]{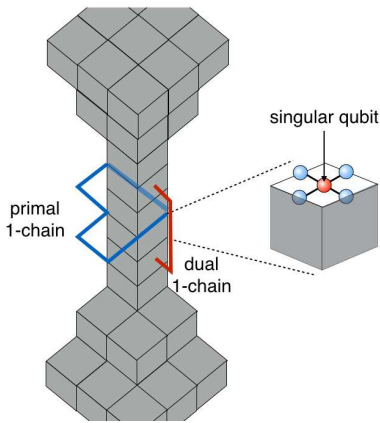}
\caption{Magic state injection without topological protection.
The primal 1-chains surrounding the defect tube result in the logical $Z$ errors on the magic state.
The dual 1-chains connecting upper and lower defect cones result in 
the logical $X$ errors.}
\label{figAp11}
\end{figure}

On the RHG lattice,
a magic state is injected by
measuring a singular qubit 
in the eigenbases of the operators $Y$ and $(X+Y)/\sqrt{2}$. 
In order to inject the magic state,
the defect is shrunk around the singular qubit
as shown in Fig.~\ref{figAp11}.
Thus the code distance around the singular qubit 
is relatively small. This causes low weight errors.
This is the reason why the singular qubit is said not to be 
topologically protected.

There are two-types of errors: one corresponds 
to primal 1-chains surrounding the shrunk defect tube and occurs as the $Z$ errors on the injected magic state (shown by a blue chain in Fig.~\ref{figAp11}),
and another corresponds to dual 1-chains connecting upper and lower 
sides of the defect cones and occurs as the $X$ errors on the injected magic state
(shown by a red chain in Fig.~\ref{figAp11}).
In order to evaluate these error accumulations,
we count the number of self-avoiding walks 
satisfying the above conditions up to length 14.
Two authors independently have built the codes for the brute force counting
and have verified to obtain the same results.
The numbers of the primal and dual 1-chains 
are listed in Table~\ref{TableSAW}.
\begin{table}[t]
\caption{The numbers of self-avoiding walks.}
\begin{center}
\begin{tabular}{c||c|c}
\hline 
length & primal & dual
\\
\hline \hline
1 & 1 & 0 
\\
\hline
2 & 0 & 0 
\\
\hline
3 & 0 & 4 
\\
\hline
4 & 7 & 8 
\\
\hline
5 & 0 & 52 
\\
\hline 
6 & 106 & 200 
\\
\hline
7 & 0 & 1060 
\\
\hline
8 & 1520 & 4084 
\\
\hline
9 & 0 & 23128 
\\
\hline
10 & 24220 & 90636 
\\
\hline
11 & 0 & 507936 
\\
\hline
12 & 409208 & 2039320 
\\
\hline
13 & 0 & 11220284 
\\
\hline
14 & 7165474 & 45854572 
\\
\hline 
\end{tabular}
\end{center}
\label{TableSAW}
\end{table}%


\end{document}